\documentclass[aps,twocolumn,floatfix,superscriptaddress]{revtex4-1}
\usepackage{amsmath,amssymb,graphicx,cases}
\usepackage{epsfig}
\usepackage{textcomp}
\usepackage{epstopdf}
\usepackage{float}
\usepackage{braket}
\usepackage[normalem]{ulem}

\newcommand{\stkout}[1]{\ifmmode\text{\sout{\ensuremath{#1}}}\else\sout{#1}\fi}

\usepackage[colorlinks=true, urlcolor=blue, anchorcolor=blue, citecolor=blue,filecolor=blue,linkcolor=blue,menucolor=blue]{hyperref}

\begin{document}
\title{The Airy distribution: experiment, large deviations and
	additional statistics}

\author{Tal Agranov}
\email{tal.agranov@mail.huji.ac.il}
\affiliation{Racah Institute of Physics, Hebrew University of
	Jerusalem, Jerusalem 91904, Israel}
\author{Pini Zilber}
\affiliation{Racah Institute of Physics, Hebrew University of
	Jerusalem, Jerusalem 91904, Israel}
\author{Naftali R. Smith}
\affiliation{Racah Institute of Physics, Hebrew University of
	Jerusalem, Jerusalem 91904, Israel}
\author{Tamir Admon}
\affiliation{Raymond and Beverly Sackler School of Chemistry, Tel Aviv University, Tel Aviv 6997801, Israel}
\author{Yael Roichman }
\email{roichman@tauex.tau.ac.il}
\affiliation{Raymond and Beverly Sackler School of Chemistry, Tel Aviv University, Tel Aviv 6997801, Israel}
\affiliation{Raymond and Beverly Sackler School of Physics and Astronomy, Tel Aviv University, Tel Aviv 6997801, Israel}
\author{Baruch Meerson }
\email{meerson@mail.huji.ac.il}
\affiliation{Racah Institute of Physics, Hebrew University of
	Jerusalem, Jerusalem 91904, Israel}

\begin{abstract}
The Airy distribution (AD) describes the probability distribution of the
area under a Brownian excursion. The AD is prominent in several
areas of physics, mathematics and computer science. Here we use a dilute colloidal
system to directly measure, for the first time, the AD in experiment. We also show how two different techniques of theory of large deviations --
the Donsker-Varadhan formalism and the optimal fluctuation method -- manifest themselves in the AD. We advance the theory of the AD by calculating, at large and small areas,  the position distribution of a Brownian excursion  conditioned on a given area, and measure its mean in the experiment. For large areas, we uncover two singularities in the large deviation function, which can be interpreted as dynamical phase transitions of third order. For small areas the position distribution coincides with the Ferrari-–Spohn
distribution, and we identify the reason for this coincidence.
\end{abstract}

\maketitle

Brownian motion came to prominence in physics and other sciences with the theoretical works of Einstein \cite{Einstein}, Smoluchowski \cite{Smoluchowski} and Langevin \cite{Langevin}, and the experimental work of Perrin \cite{Perrin}. Today, more than a hundred years since those remarkable discoveries, Brownian motion is a central paradigm in a multitude of fields \cite{satyacomputerreveiw,Mazo}. Here we focus on  some interesting properties of \emph{conditioned} Brownian motions as described by the Airy distribution and by its extensions that we will introduce.

Since its discovery nearly four decades ago \cite{dar,Louch}, the Airy distribution (AD) keeps reappearing in
seemingly unrelated problems in different fields of science.
One of its first applications was to inventory problems where the AD describes, for example, the distribution of the time spent by locomotives in a
railway
depot \cite{tac,tac2}.  The AD appears in the description of the computational cost of data storage algorithms \cite{comp1}.  In graph theory the AD is the distribution of the internal length of a rooted planar tree \cite{tac}. More recently the AD appeared in physics: as the distribution of the maximal height of fluctuating interfaces \cite{satyaprl,satcomt,solid}, the avalanche size distribution in sandpile models \cite{sand}, the size fluctuations of ring polymers \cite{poly}, and the position distribution of laser cooled atoms \cite{laser}. See Ref. \cite{satyacomputerreveiw} for a review of some of these examples.

Despite its importance,  the AD has not yet been measured in an experiment. Here we report such measurements in the simplest setting where the AD was originally discovered \cite{dar,Louch}:  the area under a Brownian excursion $x(t)$ in one dimension. We also advance the theory of, and experiments on, the AD by focusing on its previously unnoticed large-deviation properties. We show how two different large-deviation formalisms manifest themselves in the AD. This allows us to probe a previously inaccessible important quantity: the \emph{position} distribution of a Brownian excursion conditioned on a specified area.

\begin{figure}[h]
	\begin{tabular}{ll}
		\includegraphics[width=0.42\textwidth,clip=]{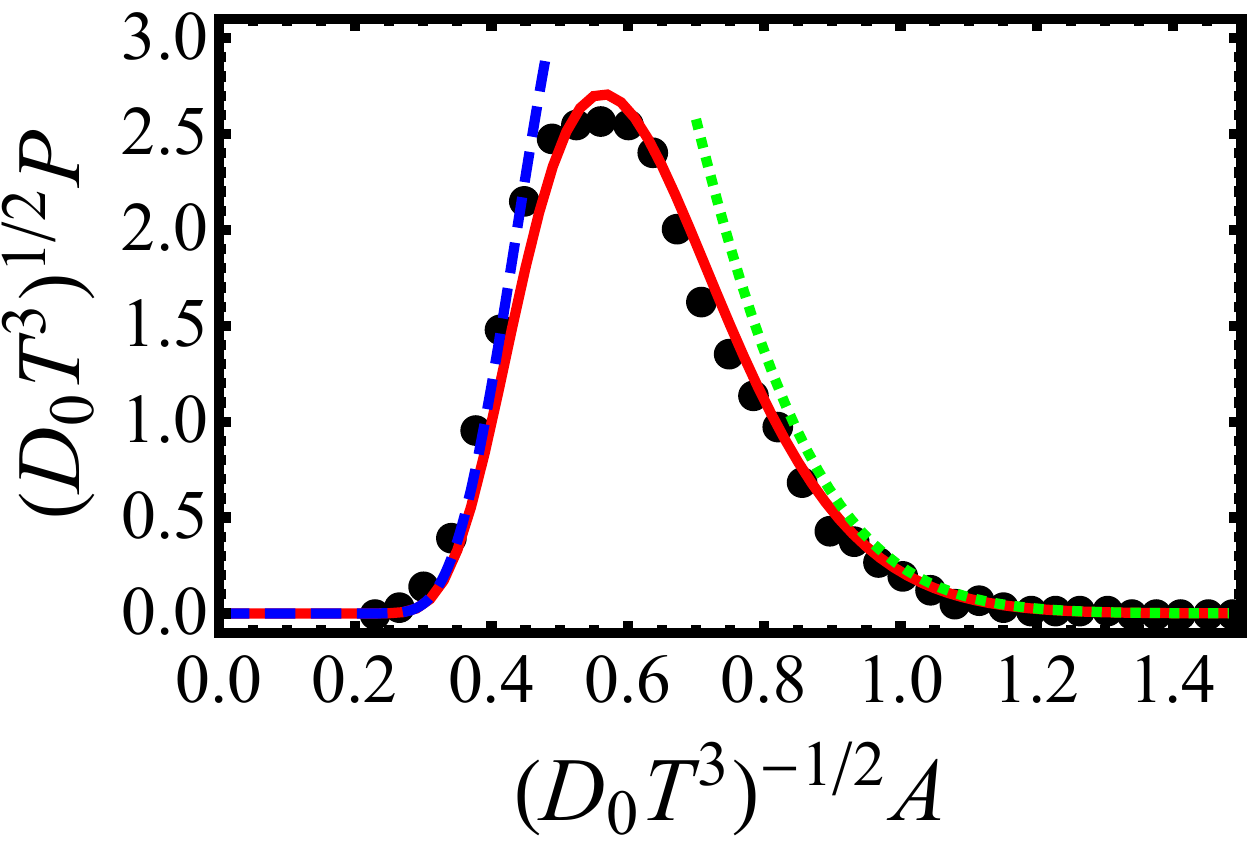}
	\end{tabular}	
\caption{Solid line: $f(\xi)$ from Eq.~(\ref{airy}). Symbols: experimentally measured histogram of the area under 12,240 excursions of duration $T\simeq 33.3$ sec. Dashed and dotted lines: the small- and large-area tails~\eqref{low} and~\eqref{high}, respectively, but with account of pre-exponential corrections, calculated in Ref. \cite{svante}.}
\label{airyfig}		
\end{figure}

Consider a Brownian excursion: a Brownian motion $x(t)$, conditioned to start and end at the origin, $x(t=0)=x(t=T)=0$, and to stay positive, $x(t)>0$ for $0<t<T$.
The area under the Brownian excursion,
\begin{equation}
A=\int_0^Tx(t)dt,\label{a}
\end{equation}
is a random variable
characterized by the probability distribution $P(A,T)$:  the AD.
The only dimensional parameters entering the problem are  $A$, $T$ and the particle diffusivity $D_0$ \cite{dif}, and dimensional analysis yields
\begin{equation}
P(A,T)=\frac{1}{\sqrt{D_0T^3}}f\left(\frac{A}{\sqrt{D_0T^3}}\right).\label{scale}
\end{equation}
The Laplace transform of the scaling function $f$, found by probabilistic methods \cite{dar,Louch}, was formally inverted to give the closed analytic form \cite{tac}
\begin{equation}
 f\left(\xi\right)=
 \frac{2\sqrt{6}}{\xi^{10/3}}\sum_{k=1}^{\infty}e^{-\beta_k/\xi^2}\beta_k^{2/3}
 U\left(-5/6,4/3,\beta_k/\xi^2\right),
 \label{airy}
 \end{equation}
where $U(\dots)$ is the confluent hypergeometric function \cite{hyper}, $\beta_k=2\alpha_k^3/27$, and $\alpha_k$ are the ordered absolute values of the zeros of the Airy function $\text{Ai}(\xi)$ \cite{airy}. The Laplace transform of $f$ was also obtained by using path integral techniques \cite{satcomt}.
The function $f(\xi)$ is shown in Fig. \ref{airyfig}.

The expression (\ref{airy}) is quite complicated. As a result, calculating the moments of the AD is already challenging \cite{satyamoment}. Here we will focus on the AD's tails \cite{cso,flajo}:
\begin{numcases}
{\!\!-\ln{P\left(A,T\right)}\simeq}
 \frac{2\alpha_1^3}{27}\frac{D_0T^3}{A^2}, & $A\ll \sqrt{D_0T^3}$, \label{low}\\
 \frac{6A^2}{D_0T^3},& $A\gg \sqrt{D_0T^3}$ , \label{high}
\end{numcases}
which are depicted, with account of pre-exponential factors, calculated in Ref. \cite{svante},  in Fig. \ref{airyfig}. They correspond to very small or very large values of the dimensionless parameter $\tilde{A}\equiv AD_0^{-1/2}T^{-3/2}$ which plays a key role in this paper. As we show here, these tails are intimately connected with two different \emph{large-deviation} formalisms of statistical mechanics. These formalisms provide a previously lacking physical picture behind these tails by characterizing the most relevant particle trajectories, which we were also able to measure in the experiment.

Our experimental setup is made of colloidal suspensions of silica spheres in water (1.50$\pm 0.08$ $\mu$m in diameter, mass density of $2.0$ g/cm$^3$, Polysciences Lot \# A762412), which are loaded into a sample cell of dimensions $22\times 22 \times 0.04$ mm constructed from a microscope slide and a cover-slip. The particles are then allowed to sediment and equilibrate and diffuse close to the bottom slide for 30 minutes at room temperature before measurements start. Quasi-2D monolayers of area fraction $\phi=0.069\pm0.005$ and $\phi=0.062\pm0.005$ are prepared by diluting the original suspension with double distilled water (DDW, 18~M$\Omega$). Sample walls are coated with bovine serum albumin to avoid particle attachment to the bottom wall of the cell.
Particle position and motion in the plane perpendicular to the optical axis are observed using bright field  microscopy (Olympus IX71). Images are captured by a CMOS camera (Grasshopper 3, Point Grey Research) at a rate of $30$~fps to allow for easy particle tracking. Conventional single particle tracking techniques were used to extract particle location with an accuracy of 6 nm \cite{Crocker1996}.
The particle diffusivity was evaluated from the particle's mean square displacement $D_0=\braket{\Delta x^2\left(t\right)}/(2t)$ averaged over all particles in the ensemble.  Excursions are constructed from the trajectories using the Vervaat transform, see  Appendix \ref{ver} for details.

Figure~\ref{airyfig} presents the measured histogram of the area under excursions of the colloidal particles. It shows a good agreement with Eqs.~\eqref{scale} and \eqref{airy}.
The histogram is a bit broader than the theoretical distribution in the region of the maximum. This effect is explained by the small but finite polydispersity of the particle diameters, leading to small variations of their diffusivity.

To start our large deviations analysis, let us define the rescaled area $a=A/T$, which is the time-averaged position of the Brownian excursion. At fixed $a$, the small-$A$ limit
(\ref{low}) corresponds to very long times, $T\gg a^2/D_0$, whereas the large-$A$ limit (\ref{high}) corresponds to very short times, $T\ll a^2/D_0$.

\begin{figure}[h]
\includegraphics[width=0.40\textwidth,clip=]{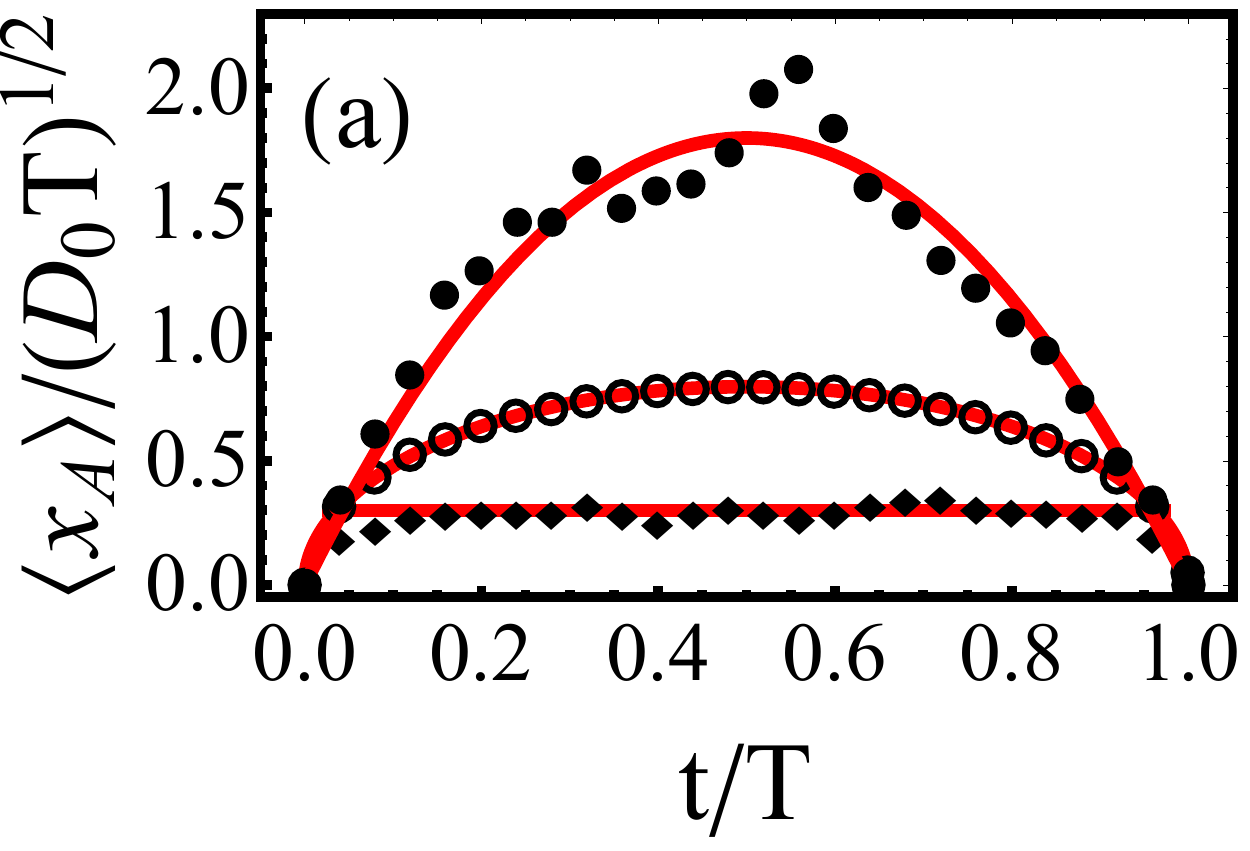}
\includegraphics[width=0.41\textwidth,clip=]{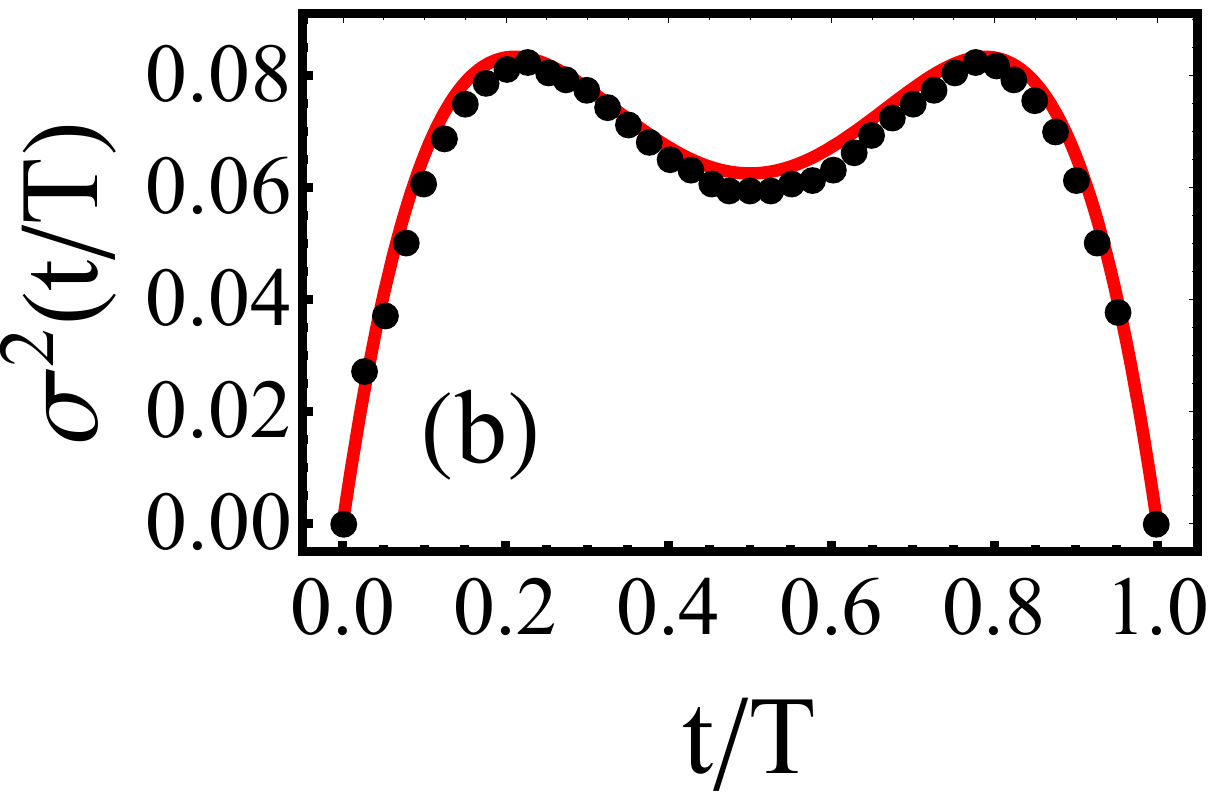}
\caption{(a) Rescaled position averages of the experimentally measured excursions, conditioned on a moderately large area,  $\tilde{A}=1.2\pm0.05$ (black circles), and small area, $\tilde{A}=0.3\pm0.05$ (black diamonds). The top and bottom lines are theoretical predictions: Eq.~\eqref{traj} with $\tilde{A}=1.2$ for large area, and the constant $\braket{x_A}/\left(D_0T\right)^{1/2}\simeq\tilde{A}=0.3$  for small area. Also shown is the rescaled average position of \emph{all} (that is, unconstrained) excursions (the empty circles) and the theoretical prediction \cite{meanex} $\braket{x(t)}/\left(D_0T\right)^{1/2}=\sqrt{8/\pi}\left[\left(t/T\right)\left(1-t/T\right)\right]^{1/2}$ (the middle line). Out of a total of $12,240$ unconditioned excursions, there were $22$ excursions within the large-area window, and $200$ within the small area window. (b) Rescaled variance of simulated excursions, conditioned on a moderately large area,  $\tilde{A}=1.2\pm0.0075$ (black circles). The line is the theoretical prediction \eqref{varhigh}. An accurate estimation of the variance requires a very large number of trajectories. These were easier to generate in simulations than in experiment.}
	\label{meantrajcomb}		
\end{figure}

Let us start with the small $A$ (or long time) limit.
We argue that the distribution of $a$ obeys a large deviation principle due to Donsker and Varadhan (DV) \cite{DonskerVaradhan,Ellis,hugo2009,Touchette2018}, where the long-time probability of observing any finite $a$ decays exponentially with time:
\begin{equation}
-\ln{ P\left(a\ll \sqrt{D_0T}\right)}\simeq T I\left(a\right).\label{dv}
\end{equation}
From dimensional analysis, the rate function $I(a)$ scales as $D_0/a^2$,
already reproducing the correct scaling behavior \eqref{low} of the small-$A$ tail. Reproducing the numerical factor $2\alpha_1^3/27$ takes slightly more effort. It boils down to determining the ground state energy of a Schr\"{o}dinger-type ``tilted operator", obtained from the generator of the constrained Brownian excursion \cite{DonskerVaradhan,Ellis}. We present this calculation in detail in Appendix \ref{dvtheo}. The trajectories $x\left(t\right)$, which mostly contribute to the small-$A$ tail (\ref{low}), stay close to the origin, without ever crossing it, for a long time. As we show below, the \emph{position distribution}, characterizing these trajectories, is stationary for most of the time, which explains the exponential decay of $P$ with time, see Eq.~\eqref{dv}. Figure~\ref{meantrajcomb}a presents the experimentally measured average position of trajectories with small $A$, which is almost constant in time as expected.

The large-$A$ tail (\ref{high}) is markedly different. It is dominated by
a \emph{single}, most probable excursion which realizes the prescribed large $A$ by straying far away from the origin during a very short time.
Other trajectories with the same $A$ have exponentially smaller probabilities. This optimal trajectory, $x_A^*(t)$, can be found by the optimal fluctuation method (OFM). For the Brownian motion, the OFM becomes \emph{geometrical optics} \cite{GF,Ikeda2015,Holcman,Meerson2019,SmithMeerson2019a,SmithMeerson2019b,3short}. The starting point of the OFM  is the path probability measure of a Brownian trajectory $x(t)$. It is given, up to pre-exponential factors, by the Wiener's action, $\mathcal P\left[x\left(t\right)\right]\propto \exp \left(-s/D_0\right)$, where
 \begin{equation}
 s\left[x\left(t\right)\right]=\frac{1}{2}\int_0^T dt\,\dot{x}^2(t).\label{s1}
 \end{equation}
The optimal trajectory can be found by minimizing the action (\ref{s1}) along excursions $x\left(t\right)$ subject to the constraint (\ref{a}). The latter can be accommodated via a Lagrange multiplier $\lambda$, leading to an effective Lagrangian  $L\left(x, \dot{x}\right) = \dot{x}^2/2-\lambda x$.
The optimal trajectory is  a parabola,
 \begin{equation}
 x_A^*\left(t\right)= (6At/T^2)\left(1-t/T\right), \label{traj}
 \end{equation}
where we have imposed $x(0)=x(T)=0$ and set $\lambda=12A/T^3$ to obey Eq.~\eqref{a}. $x_A^*\left(t\right)$ is also the average position over all trajectories with the large prescribed $A$. Figure~\ref{meantrajcomb}a compares  Eq.~\eqref{traj} with the experimentally measured average position of excursions with large $A$. Although the OFM becomes asymptotically exact only in the limit of $\tilde{A} \gg 1$, a good agreement is observed already for $\tilde{A} \gtrsim 1$.

Plugging Eq.~(\ref{traj}) into the action \eqref{s1}, we exactly reproduce
Eq.~\eqref{high}, see also Refs. \cite{FillJanson,3short}. The large-$A$ tail~\eqref{high} is shared by other Brownian motions (such as the Brownian \emph{bridge} and its absolute value \cite{jon})  which start at $x=0$ at $t=0$ and return to $x=0$ at $t=T$, but are allowed to reach or even cross $x=0$ at $0<t<T$.  The large-$A$ tails coincide because the optimal trajectory \eqref{traj} is unaffected by the non-crossing constraint.

\begin{figure}[h]
		\includegraphics[width=0.40\textwidth,clip=]{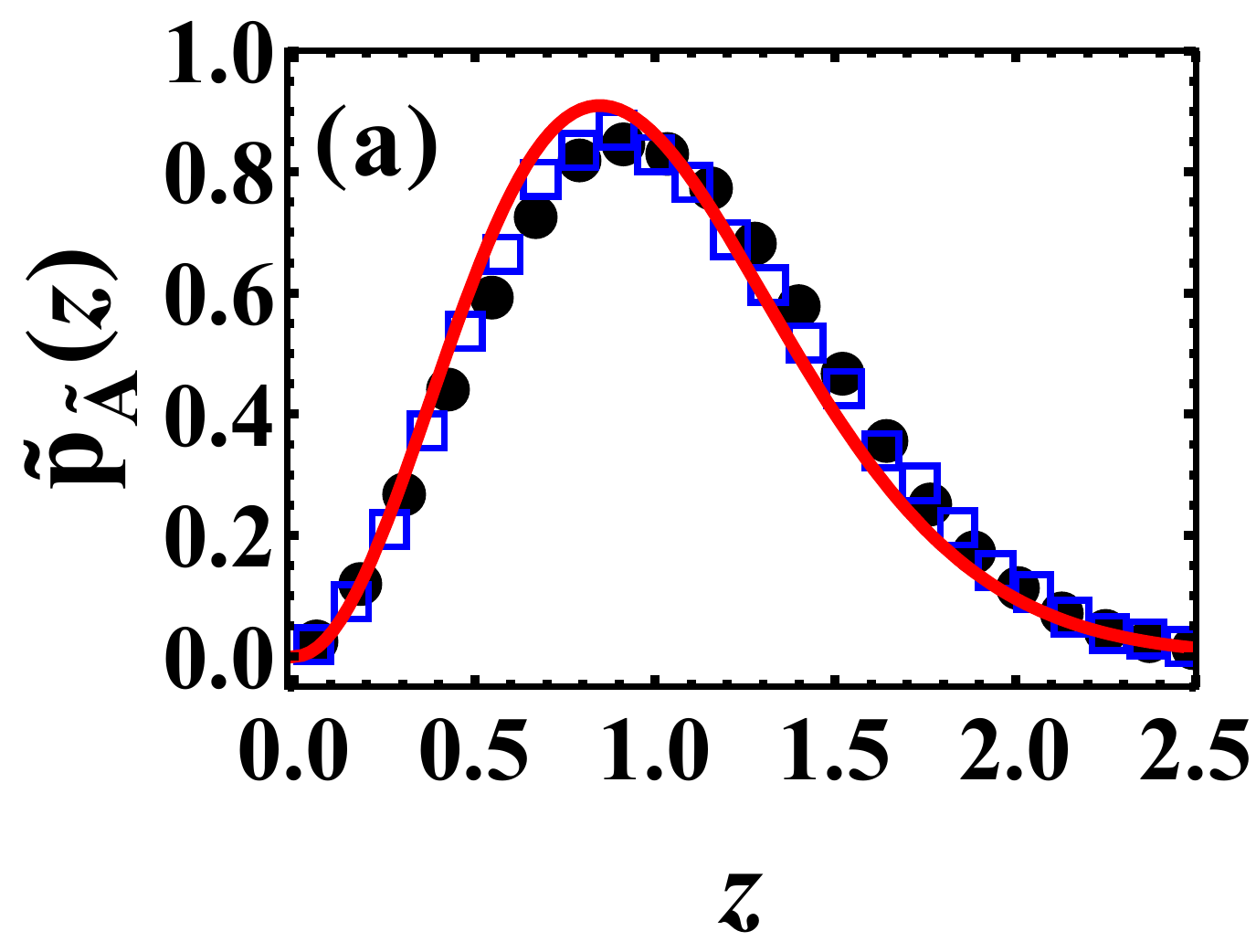}
		\includegraphics[width=0.38\textwidth,clip=]{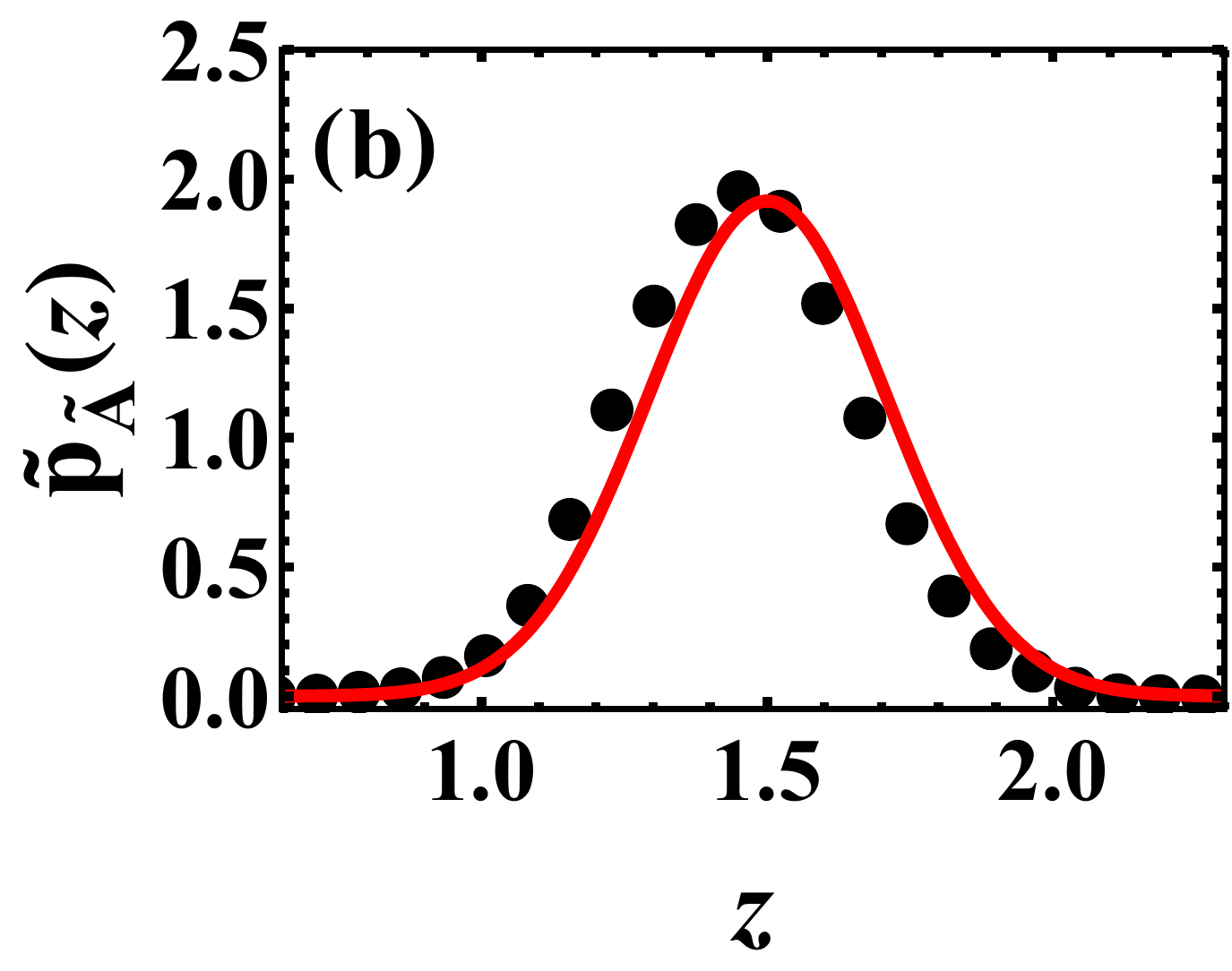}
	\caption{Black circles: the position histograms $\tilde{p}_{\tilde{A}}\left(z,1/2\right)$ of simulated excursions conditioned on $A$ for  small area $\tilde{A}=0.3\pm0.0075$ (a) and large area $\tilde{A}=1.2\pm0.0075$  (b). Lines:  theoretical predictions \eqref{condlow} (a) and \eqref{ldf} and \eqref{fsub} (b). In the latter prediction the normalization factor was accounted for. Panel (a) also shows the histogram at $t/T=3/4$ (squares), confirming the time-independence of the position distribution at small areas.}
	\label{condfig}		
\end{figure}

A more detailed characterization of the trajectories with specified area is given by the \emph{position} distribution of the excursion, $p\left[x_A(t)=X|A,T\right]\equiv p_A\left(X,t\right)$, conditioned on a given area $\int_0^Tx_A(t)dt=A$. This important distribution has been previously inaccessible \cite{inaccess}. As we show now, the two large-deviation techniques  that describe the two tails of $f(\xi)$, allow one to  evaluate $p_A$ at small and large $A$.
From dimensional analysis,  the distribution must have the scaling form
\begin{equation}
p_A\left(X,t\right)=\frac{T}{A}\tilde{p}_{\tilde{A}}\left(\frac{XT}{A},\frac{t}{T}\right).
\end{equation}
At $\tilde{A}\ll 1$, the conditional distribution can be found with the DV formalism \cite{dvconditioned}, see Appendix \ref{dvtheo}. Apart from narrow temporal boundary layers at $t=0$ and $t=T$, this formalism predicts the stationary position distribution
\begin{equation}
\tilde{p}_{\tilde{A}}\left(z,\frac{t}{T}\right)\simeq \frac{2\alpha_1 }{3}
\frac{\text{Ai}^2\left(\frac{2\alpha_1}{3}z-\alpha_1\right)}
{\left[\text{Ai}^{\prime}\left(-\alpha_1\right)\right]^2},\quad \tilde{A}\ll 1,\label{condlow}
\end{equation}
where $\text{Ai}^{\prime}(z)\equiv (d/dz)\text{Ai}(z)$.
The first moment of the distribution (\ref{condlow}) is equal to unity, which gives $\braket{x_A\left(t\right)}\simeq A/T$ in the dimensional variables. Figure \ref{condfig}a shows good agreement between the distribution of simulated excursions, conditioned by small area, and Eq.~\eqref{condlow}. Simulation details are described in Appendix \ref{ver}.  See also Fig. \ref{meantrajcomb}a for a comparison of the mean which we measured in the experiment.

The distribution \eqref{condlow} is known as the Ferrari-Spohn (FS) distribution.  It first appeared in the studies of position fluctuations of a Brownian excursion, conditioned to stay away from a wall $x_w(t)$ which moves, sufficiently fast, back and forth so that $x_w(0)=x_w(T)=0$ and $x_w(0<t<T)>0$
\cite{Groeneboom,FM2000,fer}. For a parabolic $x_w(t)$,
\begin{equation}\label{wall}
x_w(t)=Ct\left(1-t/T\right),
\end{equation}
the small-$A$ distribution~(\ref{condlow}) and the FS distribution \cite{fer} \emph{coincide}. This unexpected coincidence can be explained by an exact mapping that we found between the two systems. The details of calculation are presented in  Appendix \ref{mapping}. The mapping involves a \emph{biased} ensemble \cite{hugoprl}:  an ensemble of excursions, where the probability of each excursion is re-weighted by the exponential factor $e^{-\mu A\left[x\left(t\right)\right]}$, where $A\left[x\left(t\right)\right]$ is the area of the excursion. This ensemble  and the ensemble of excursions conditioned on $A$ are equivalent, and they obey the same relation as the one between the canonical and micro-canonical ensembles, respectively, in equilibrium statistical mechanics \cite{hugoprl}. If we denote by $p\left(X,t;\mu\right)$ the position distribution in the biased ensemble, and by $p_C\left(X,t\right)$ the position distribution of the particle relative to the parabolic wall~(\ref{wall}) in the FS problem, the mapping reads $p\left(X,t;\mu=2C/D_0T\right)=p_C\left(X,t\right)$,  see Appendix \ref{mapping}. The FS distribution has also appeared in other systems \cite{fs1,fs2,fs3,nach,SmithMeerson2019b} which is indicative of a universality class.

\begin{figure}[h]
		\includegraphics[width=0.42\textwidth,clip=]{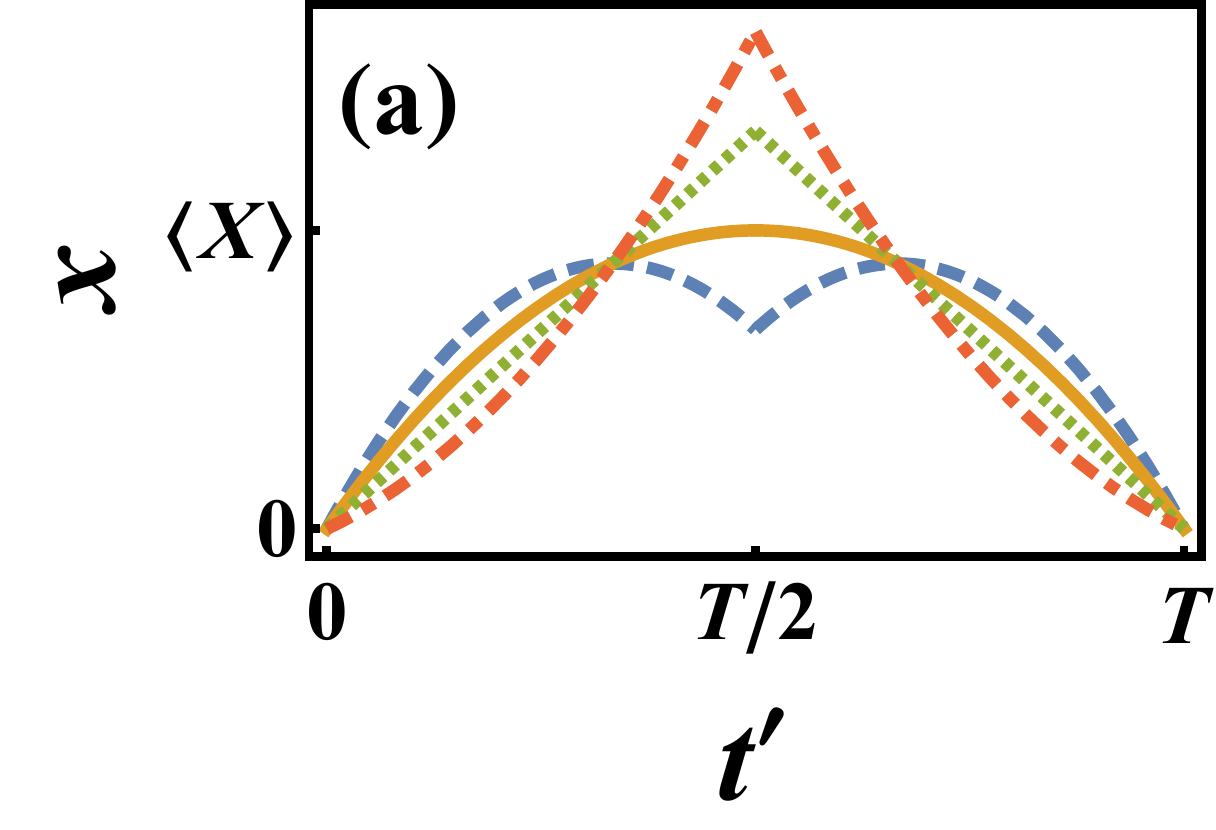}
		
		\includegraphics[width=0.42\textwidth,clip=]{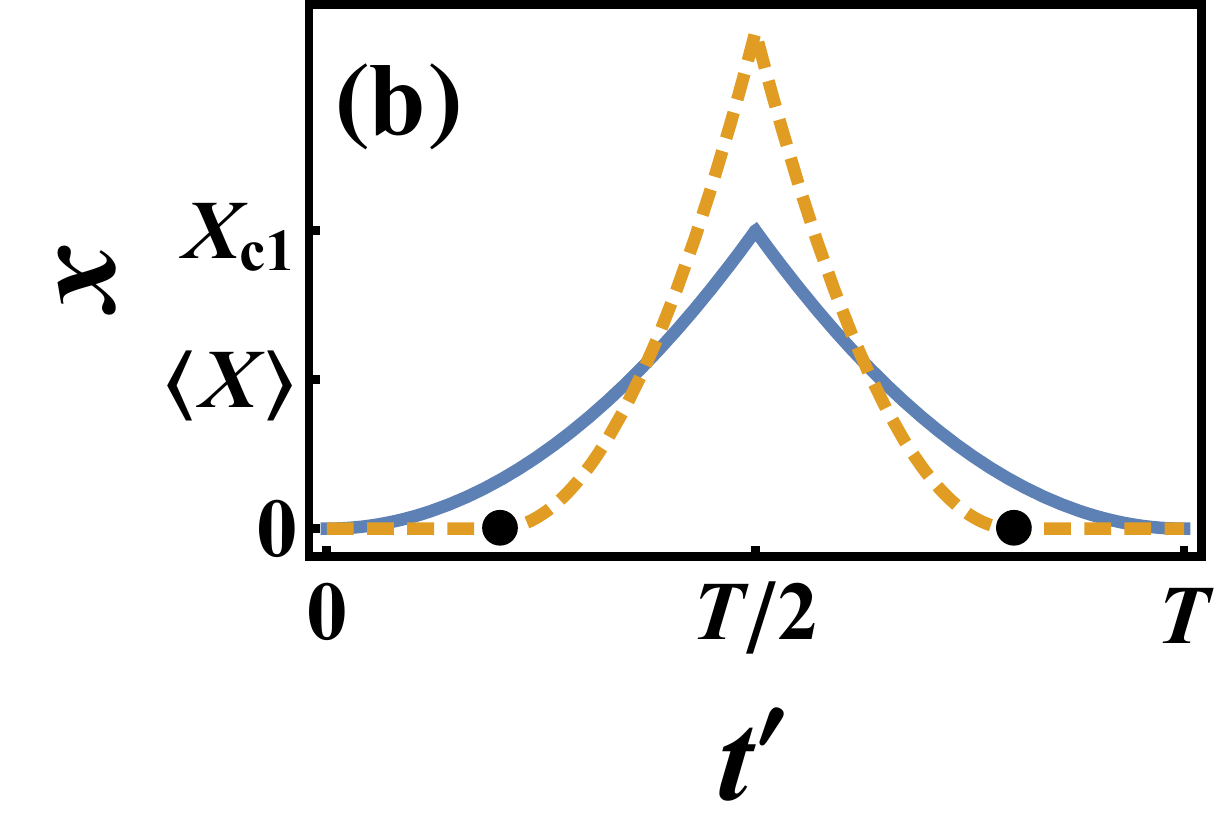}
	\caption{Optimal paths conditioned on $x\left(t=T/2\right)=X$ and the area $A$, in the subcritical (a) and supercritical (b) regimes, see Eqs.~\eqref{trajsub} and~\eqref{trajsup}. The parameters are $XT/A=1,3/2,2$ and $5/2$ (dashed, solid, dotted and dot-dashed lines, respectively) (a) and $XT/A=3$ and $5$ (solid and dashed lines, respectively) (b). Circles: the points $ |T/2-t^{\prime}|= 3A/\left(2X\right)$.}
	\label{wkbtraj}		
\end{figure}

At $\tilde{A}\gg 1$ the conditional distribution $\tilde{p}_{\tilde{A}}\left(z,t/T\right)$ is time-dependent, and it can be found with the OFM, see Appendix \ref{ofm}. The probability of an excursion to reach a specified position $x\left(t\right)=X$ \emph{and} to have a given large area $A$ comes from the optimal trajectory which minimizes the action~\eqref{s1} under the two constraints. The  optimal trajectory consists of two parabolic segments, joined at time $t$ with a corner singularity there, which originates from conditioning on the specified position at time $t$, see Appendix \ref{ofm}. For the particular case of conditioning on the position $x=X$ at
$t=T/2$, the optimal trajectory is
\begin{equation}
x(t^{\prime}) =  X+\left(6a-4X\right)\left|1-\frac{2t^{\prime}}{T}\right|
-\left(6a-3X\right)\left(1-\frac{2t^{\prime}}{T}\right)^2 \label{trajsub}
 \end{equation}
(recall that $a=A/T$). This trajectory is shown, for several values of $X$, in Fig. \ref{wkbtraj}a.
However, the solution~(\ref{trajsub}) is a legitimate Brownian excursion only when $X$ is smaller than a critical value $X_{c1}=3a$.
For $X>3a, $ $x(t^{\prime})$ from Eq.~(\ref{trajsub}) would cross the origin, which is forbidden.  The correct solution for $X>3a$ is provided by the \emph{tangent construction} of the calculus of one-sided variations \cite{onesided}, and we obtain
\begin{equation}
\label{trajsup}
x(t^{\prime}) =
\begin{cases}
\frac{4X^3}{9A^2}\left(\frac{3A}{2X}-\left|\frac{T}{2}-t^{\prime}\right|\right)^2, & |\frac{T}{2}-t^{\prime}|\leq \frac{3A}{2X},\\
0,                           & \left|\frac{T}{2}-t^{\prime}\right|\geq \frac{3A}{2X},
\end{cases}
\end{equation}
see Fig. \ref{wkbtraj}b.  The conditional distribution is given by $\Delta s$: the difference between the action \eqref{s1} along the additionally constrained trajectory, Eq. \eqref{trajsub} or \eqref{trajsup}, and the action along the optimal trajectory  \eqref{traj} constrained by the area alone. As a result,
\begin{equation}
-\ln \tilde{p}_{\tilde{A}}\left(z,\frac{1}{2}\right) \simeq \frac{\Delta s}{D_0} = \frac{A^2}{D_0T^3}g\left(z\right),\quad\tilde{A}\gg 1.\label{ldf}
\end{equation}
The large deviation function
\begin{numcases}
{\!\!g\left(z\right)=}
8\left(z-3/2\right)^2, & $0<z\leq 3$, \label{fsub}\\
(8/9)z^3 -6, & $z\geq 3$, \label{fsup}
\end{numcases}
has a singularity: its third derivative is discontinuous at $z=3$. This singularity can be interpreted as a dynamical phase transition of the third order. Similar singularities have been recently found in other Brownian motions pushed into large-deviation regimes by constraints \cite{Meerson2019,SmithMeerson2019a,SmithMeerson2019b,3short}. A sharp transition, however, appears only in the limit of $\tilde{A}\to \infty$, and is smoothed out at finite $\tilde{A}$.

The sub-critical result \eqref{fsub} describes Gaussian fluctuations around the mean
value $\bar{z}=3/2$ which corresponds to an excursion conditioned on $A$ but not on $X$. This result is in good agreement with simulations, see Fig.~\ref{condfig}b. In the supercritical regime, Eq.~\eqref{fsup},
the support of the optimal trajectory is shorter than $T$, see Fig. \ref{wkbtraj}b. As a result, the time $T$ cannot enter the final result. This, and the scaling relation~\eqref{ldf}, explain the fact that the first term in Eq.~\eqref{fsup} [which comes from the action along the supercritical trajectory \eqref{trajsup}] is cubic.

If the condition $x\left(t\right)=X$ is specified at $t \neq T/2$, the solution becomes a bit more involved; we present it in Appendix \ref{ofm}. Here the third-order dynamical phase transition occurs at a \emph{critical line} $X_{c1}\left(t\right)$. Furthermore, the optimal trajectory becomes asymmetric around $T/2$, and an additional third-order transition occurs at a higher critical value $X_{c2}\left(t\right)>X_{c1}\left(t\right)$, similarly to the third-order transition, recently found in a different constrained Brownian motion \cite{SmithMeerson2019a}.  In the subcritical regime, $X\leq X_{c1}\left(t\right)$, Eqs.~(\ref{ldf}) and~\eqref{fsub} give way to
\begin{equation}
-\ln \tilde{p}_{\tilde{A}}\left(z,\frac{t}{T}\right) \simeq
\frac{A^2}{D_0T^3}\frac{\left[z-\bar{z}\left(t\right)\right]^2}{2\sigma^2\left(t/T\right)},\label{ldf_t}
\end{equation}	
where $\bar{z}\left(t\right)= Tx_A^*\left(t\right)/A$, see Eq.~\eqref{traj}, and
\begin{equation}
\sigma^2\left(\xi\right)=\xi \left(1-\xi\right)\left(3 \xi^2-3\xi+1\right).\label{varhigh}
\end{equation}
These predictions also agree with simulations, see Fig.~\ref{meantrajcomb} b. As the excursions start and end at $x=0$, the variance vanishes at $t=0$ and $t=T$. More surprisingly, the variance \eqref{varhigh} has a local minimum at $t=T/2$ and is maximal at $t=T\left(3\pm\sqrt{3}\right)/6$. The appearance of the local minimum of $\sigma^2(t/T)$ at $t=T/2$ is not exclusive to the large-$A$ limit: we observed it in simulations for all values of $\tilde{A}$, but it is most prominent in the large-$A$ tail.
\\
As shown in Appendix \ref{ofm}, in the subcritical regime $X\leq X_{c1}\left(t\right)$, the optimal trajectory is unaffected by the constraint at the origin for any $t$. This explains the coincidence of the Gaussian fluctuations in this regime, Eqs.~\eqref{ldf_t} and~\eqref{varhigh}, with those of a Brownian \emph{bridge} conditioned on $A$ \cite{mazollo,mazollo2}.

The two large deviation formalisms -- the DV method and the OFM -- can be applied to other stochastic processes and dynamical observables. One example is the distribution $P(B,T)$ of the area under the \emph{square} of a Brownian excursion, $B = \int_0^T x^2(t) dt$. This distribution exhibits the scaling behavior $P(B,T) = D_0^{-1}T^{-2}h\left(B/D_0T^2\right)$. We show in Appendix \ref{square} how one can use the DV method and the OFM to obtain
the \emph{tails} of the scaling function $h(\dots)$, which were previously derived by probabilistic methods \cite{cso}.

In conclusion, we presented the first direct experimental measurements of the Airy distribution (AD) and of some of its large deviation extensions. By exploiting the connection with two different large-deviation formalisms, we uncovered a relation of the AD with the Ferrari-Spohn distribution and found two dynamical phase transitions.

A promising future direction is to study the statistics of time-integrated quantities in
\emph{multi-particle} systems. For a large number of particles such systems can be efficiently probed with still another large-deviation technique: the rapidly developing macroscopic fluctuation theory \cite{MFTreview}.

B.M. was supported by the Israel Science Foundation (Grant No. 807/16). N.R.S. was supported by the Clore Israel Foundation.

\appendix

\section{Experiment, simulations and the Vervaat's transform}\label{ver}

\renewcommand{\theequation}{A\arabic{equation}}
\setcounter{equation}{0}

Experiment and numerical simulations produce free Brownian trajectories, rather than Brownian excursions.
We obtained the latter from the former by applying two successive transformations, see \textit{e.g.} Ref. \cite{conditionedsatya2}.
First, we employ a well-known mapping which transforms a free Brownian motion $x_{\text{Bm}}\left(t\right)$ into a Brownian bridge $x_{\text{Br}}\left(t\right)$ of duration $T$ \cite{feller}:
\begin{equation}
	x_{\text{Br}}\left(t\right)=x_{\text{Bm}}\left(t\right)-\frac{t}{T}x_{\text{Bm}}\left(T\right).\label{a1}
\end{equation}
Next we employ the Vervaat's transform \cite{var} which transforms a Brownian bridge into a Brownian excursion $x_{\text{Ex}}\left(t\right)$. The Vervaat's transform can be realized in three steps. First, place the origin at the absolute minimum attained by the bridge, such that it is positive at all times. Next, shift the time by $\tau$, the time at which the minimum was attained: $t\rightarrow t+\tau$. Finally, glue the first part of the trajectory from $t=0$ to $t=\tau$ with the remainder of the trajectory from $t=\tau$ to $t=T$:
\begin{equation}
	x_{\text{Ex}}\left(t\right)=x_{\text{Br}}\left(t+\tau \mod{T}\right)-x_{\text{Br}}\left(\tau\right).\label{a2}
\end{equation}
This procedure yields, in experiment and simulations, an ensemble of excursions with a prescribed duration $T$.

\begin{figure}[h]
	\begin{tabular}{ll}
		\includegraphics[width=0.42\textwidth,clip=]{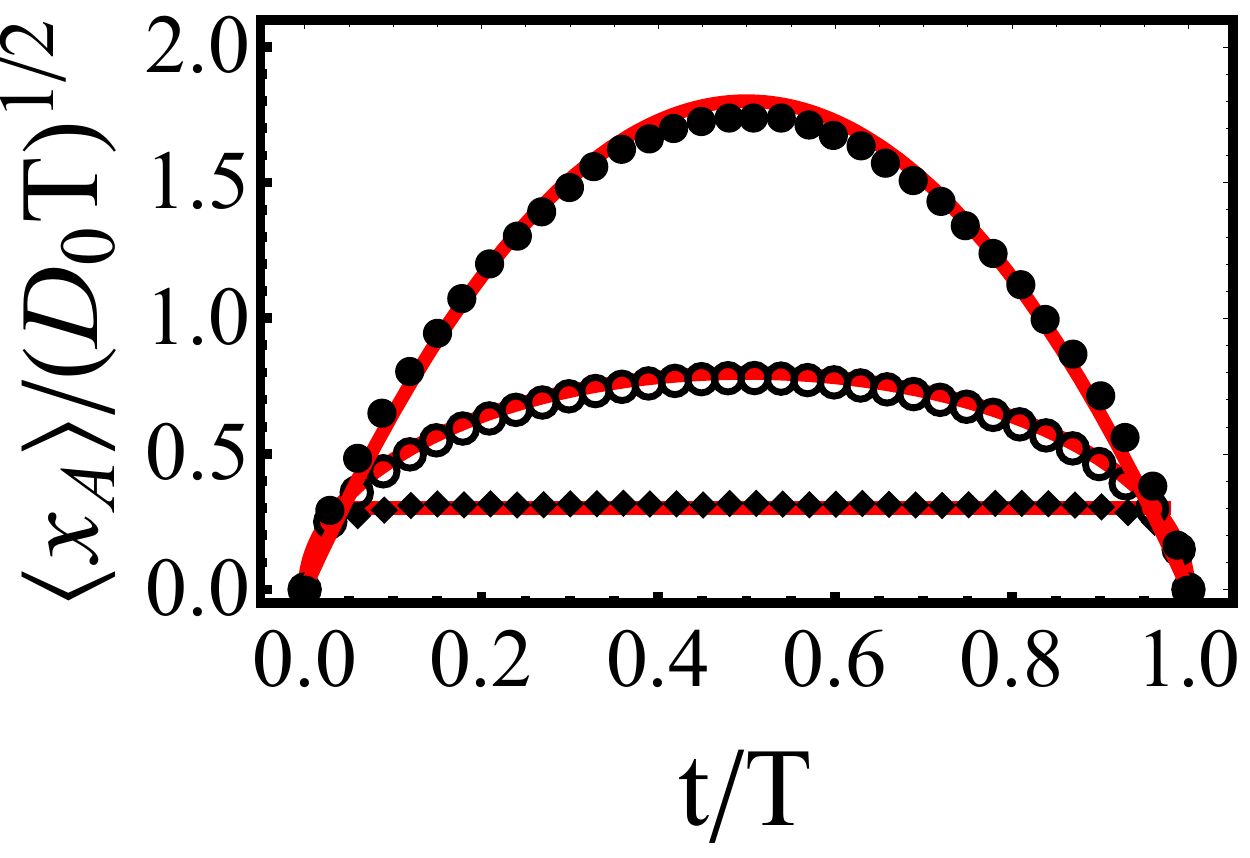}
	\end{tabular}	
	\caption{Rescaled empirical position averages over simulated excursions, conditioned on a moderately large area, $\tilde{A}=1.2\pm0.0075$ (black circles), and small area, $\tilde{A}=0.3\pm0.0075$ (black diamonds). The top and bottom lines are theoretical predictions: Eq.(8) of the main text with $\tilde{A}=1.2$ for large area, and the constant $\braket{x_A}/\left(D_0T\right)^{1/2}\simeq\tilde{A}=0.3$  for small area. Also shown is the rescaled empirical average position of \emph{all} (that is, unconstrained) excursions (the empty circles) and the theoretical prediction \cite{meanex} $\braket{x(t)}/\left(D_0T\right)^{1/2}=\sqrt{8/\pi}\left[\left(t/T\right)\left(1-t/T\right)\right]^{1/2}$ (the middle line).}
	\label{meantrajcombsimo}		
\end{figure}

The mean position of excursions with large and small area $A$ was measured in the experiment and was found to agree well with the theoretical prediction as shown in Fig. 2a of the main text. The full position distribution of excursions, conditioned on a given area, requires a considerably larger number of trajectories. It was easier to meet this requirement in simulations. In our simulations with the relatively large area under excursion we used $5,324$ trajectories whose areas fit into the window $\tilde{A}=1.2\pm0.0075$. These were extracted from a total of $2\times 10^7$ Brownian excursions unconditioned by a specified area. Each trajectory was sampled along $1,000$ equally spaced points during its dynamics. For the small-area simulations, in the window  $\tilde{A}=0.3\pm0.0075$, we used $1,885$ trajectories. These were extracted from a total of $2\times10^6$ unconditioned Brownian excursions.  Each trajectory here was sampled along $5,000$ equally spaced points. We verified the simulation method by measuring the area distribution of the excursions and comparing the results with the theoretical prediction (2) and (3) of the main text. Very good agreement was observed. A very good agreement between the theory and the simulations is also evident for the mean position conditioned on high and low area $A$, as shown in Fig.~\ref{meantrajcombsimo}. On account of the very large number of simulated trajectories, the agreement here is better than the one observed in Fig. 2a of the main text.

\section{\label{dvtheo}The DV formalism and the small-$A$ tail}

\renewcommand{\theequation}{B\arabic{equation}}
\setcounter{equation}{0}

The DV formalism \cite{DonskerVaradhan} reduces the problem of finding the rate function $I(a)$, defined in Eq.~(6) of the main text, in the large $T$ (or small $A$) limit, to an effective eigenvalue problem. According to G\"{a}rtner and Ellis \cite{Ellis}, the rate function $I(a)$ is given by a Legendre-Fenchel transform
\begin{equation}\label{GE}
	I(a) = \max_k \left[ k a - \tilde{I}(k) \right]
\end{equation}
of the scaled cumulant generating function (SCGF)
\begin{align}
	\tilde{I}(k) &= \lim_{T\to \infty} \frac{1}{T} \ln \braket{{e^{Tk a}}},  \label{SCGF}
\end{align}
where $\braket{...}$ denotes averaging over values of $a$ with respect to the Airy distribution~(2). According to the DV method,
\begin{equation}\label{maxEV}
	\tilde{I}(k) = \xi_\text{max}\left(\hat{L}^{(k)}\right),
\end{equation}
where $\xi_\text{max}$ is the maximal eigenvalue of the operator $\hat{L}^{(k)} \equiv \hat{L} + k x$, which is a tilted version of the Fokker-Planck generator $\hat{L}$, defined by the Fokker-Planck equation
\begin{equation}
	\frac{\partial P(x,t)}{\partial t} = \hat{L} P(x,t), \label{tilt}
\end{equation}
corresponding to the Langevin equation for the stochastic process in question.  For Brownian excursions the Fokker-Planck generator is just $\hat{L}=(D_0/2)\, \partial^2_{x}$
for $x>0$, and $\hat{L} = -\infty$ for $x\leq 0$.
It is convenient to convert the Fokker-Planck problem into an effective quantum mechanical one by considering the minus Fokker-Planck generator as an effective Hamiltonian, $\hat{H} \equiv -\hat{L}$, and look for ground state energy $E_\text{min}$ of the latter \cite{hugo2009}. As a result,
\begin{equation}\label{Quantum}
	\tilde{I}(k) = -E_\text{min}\left(\hat{H}^{(k)}\right) .
\end{equation}
The effective Schr\"{o}dinger equation for the Brownian excursion reads
\begin{equation}\label{Schrodinger}
	-\frac{D_0}{2}\frac{d^2\psi^{(k)}}{dx^2}-k x\psi^{(k)}=E^{(k)} \psi^{(k)},
\end{equation}
with the boundary conditions $\psi^{(k)}(0) = \psi^{(k)}(\infty) = 0$. The (discrete) spectrum, corresponding to Eq.~(\ref{Schrodinger}) with $k < 0$, is $E_n^{(k)} = \left(D_0/2\right)^{1/3}k^{2/3}\alpha_n $ where $n=1,2,\dots$, and $\alpha_n$ are the absolute values of the zeros of the Airy function
\cite{airy}. The corresponding eigenfunctions are $$\psi_n^{(k)}(x)\propto \text{Ai}\left[\left(-2k/D_0\right)^{1/3}x-\alpha_n\right] .
$$
The SCGF is given by
\begin{equation}\label{SCGF_result}
	\tilde{I}(k) = -\left(D_0/2\right)^{1/3} k^{2/3}\alpha_1.
\end{equation}
Plugging this result into Eq. (\ref{GE}) (and again assuming $k< 0$), we see that the maximum is achieved for
\begin{equation}
	k\left(a\right)=-4D_0\left(\alpha_1/3a\right)^3,\label{optlam}
\end{equation}
and Eq. (\ref{GE}) yields the small-$A$ asymptotic~(4) of the main text.

The solution of the tilted-generator problem also provides the conditional position distribution,  associated with a prescribed rescaled area $a$. This distribution is given by the product of the left and right eigenfunctions of the tilted operator $\hat{L}^{(k)}$, corresponding to the maximal eigenvalue. \cite{dvconditioned}. In our case the left and right eigenfunctions coincide due to the hermiticity of the tilted operator, and are both given by $\psi_1^{k}\left(x\right)$. Their properly normalized product gives the conditional position distribution that appears in Eq.~(10) of the main text, after the substitution of $k\left(a\right)$ from Eq.~\eqref{optlam}.

\section{Mapping to Ferrari-Spohn model}\label{mapping}
\renewcommand{\theequation}{C\arabic{equation}}
\setcounter{equation}{0}

Ferrari and Spohn (FS) \cite{fer} studied the statistics of the position, at an intermediate time $t=t'$, of a Brownian bridge $x\left(t\right)$, when the process is constrained on staying away from an absorbing wall, that is $x\left(t\right)>x_{\text{w}}(t)$, where $x_{\text{w}}(t)$ is a semicircle, $x_{\text{w}}(t)=Ct^{1/2}\left(T-t\right)^{1/2}$. They also extended their results to other concave (that is, convex upward) functions.
FS proved that at $T \to \infty$, typical fluctuations of $\Delta X\!=\!x\left(t'\right)-x_{\text{w}}\left(t'\right)$ away from the moving wall obey a universal distribution which depends only on the second derivative $\ddot{x}_{\text{w}}\left(t'\right)$. This universal distribution can be represented as
\begin{equation}
	\label{eq:FS_distribution_with_ell}
	P\left(\Delta X\right)=\frac{\ell\text{Ai}^{2}\left(\ell\Delta X-\alpha_{1}\right)}{\text{Ai}'\left(-\alpha_{1}\right)^{2}},
\end{equation}
where $\text{Ai}\left(\dots\right)$ is the Airy function, $\alpha_1=+2.338107\dots$ is the magnitude of its first zero, $\text{Ai}'$ is the derivative of the Airy function with respect to its argument and
$\ell=\left[-2\ddot{x}_{\text{w}}\left(t'\right)/D_{0}^{2}\right]^{1/3}\!$. Eq.~(\ref{eq:FS_distribution_with_ell}) is valid in the limit $CT\gg\sqrt{D_{0}T}$, that is when the wall ``pushes'' the Brownian bridge into the large-deviation regime.

Remarkably, Eq.~(10) of the main text, which describes the single-point distribution of a Brownian excursion conditioned on its covering a very small area $A\ll D_{0}^{1/2}T^{3/2}$, coincides with the distribution~(\ref{eq:FS_distribution_with_ell})  with $\ell=2\alpha_{1}/\left(3a\right)$. This suggests that the model studied in the main text is related to the FS model.
Indeed, we now establish a formal mapping between the two models, and explain this coincidence.

The path integral that corresponds to the FS model is
\begin{equation}
	\label{eq:path_integral_x}
	\int Dx\left(t\right)e^{-s\left[x\left(t\right)\right]/D_{0}},
\end{equation}
constrained by $x\left(t=0\right)=x\left(t=T\right)=0$ and $x\left(t\right)>x_{\text{w}}\left(t\right)$, where $s\left[x\left(t\right)\right]$ is the Wiener's action, given by Eq.~(7) of the main text.
Let us define $y\left(t\right)=x\left(t\right)-x_{\text{w}}\left(t\right)$. Rewriting Eq.~(\ref{eq:path_integral_x}) in terms of $y\left(t\right)$ (the Jacobian of this transformation is equal to $1$), we obtain
\begin{equation}
	\label{eq:path_integral_y}
	\int Dy\left(t\right)e^{-\tilde{s}\left[y\left(t\right)\right]/D_{0}},
\end{equation}
where $y\left(t\right)$ are Brownian excursions, $y\left(t=0\right)=y\left(t=T\right)=0$ and $y\left(t\right)>0$, and the action is

\begin{eqnarray}
	\label{eq:action_y_gen}\nonumber
	\tilde{s}\left[y\left(t\right)\right]&=&s\left[x_{\text{w}}\left(t\right)+y\left(t\right)\right]
	=s_{0}+\frac{1}{2}\int_{0}^{T}\left(\dot{y}^{2}+2\dot{x}_{\text{w}}\dot{y}\right)dt\\
	&=&s_{0}+\frac{1}{2}\int_{0}^{T}\left(\dot{y}^{2}-2\ddot{x}_{\text{w}}y\right)dt ,
\end{eqnarray}
where we used integration by parts and defined
\begin{equation}
	s_{0}=\frac{1}{2}\int_{0}^{T}\dot{x}_{\text{w}}^{2}\,dt ,
\end{equation}
which is independent of $y\left(t\right)$.
For the particular case of a parabolic wall $x_{\text{w}}\left(t\right)=Ct\left(1-t/T\right),$  we have $\ddot{x}_{\text{w}}\left(t\right)=-2C/T$, so Eq.~(\ref{eq:action_y_gen}) becomes
\begin{equation}
	\label{eq:action_y}
	\tilde{s}\left[y\left(t\right)\right]=s\left[y\left(t\right)\right]+\frac{2C}{T}\int_{0}^{T}y\left(t\right)dt+s_{0}.
\end{equation}
The distribution of $\Delta X$ is given by
\begin{equation}
	\label{eq:P_ofDeltaX_path_integral}
	P_{C}\left(\Delta X\right)=\frac{\int Dy\left(t\right)e^{-\tilde{s}\left[y\left(t\right)\right]/D_{0}}\delta\left[y\left(t'\right)-\Delta X\right]}{\int Dy\left(t\right)e^{-\tilde{s}\left[y\left(t\right)\right]/D_{0}}}
\end{equation}
where the $C$-dependence enters through $\tilde{s}$.
The constant $s_0$ is of no importance because its contributions cancel out.
Equation~(\ref{eq:P_ofDeltaX_path_integral}) is exact. In the large-deviation limit $CT\gg\sqrt{D_{0}T}$,  $P_{C}\left(\Delta X\right)$ is given by the FS distribution~(\ref{eq:FS_distribution_with_ell}) with
\begin{equation}
	\label{eq:ell_as_function_of_C}
	\ell=\left(\frac{4C}{D_{0}^{2}T}\right)^{1/3} .
\end{equation}

We now wish to find a connection between the FS model and the model studied in the present work. Let us begin by defining the canonical or biased ensemble \cite{hugoprl}. This is an ensemble of excursions which is unconstrained by a specified $A$, but where the probability of each excursion is re-weighted by the exponential factor $e^{-\mu A\left[x\left(t\right)\right]}$, where $A\left[x\left(t\right)\right]$ is the area of the excursion.
The distribution of $X$ in the canonical ensemble is given by
\begin{equation}
	\label{eq:canonical_ensemble_as_path_integral}
	p\left(X;\mu\right)=\frac{\int Dx\left(t\right)e^{-s_{\mu}\left[x\left(t\right)\right]/D_{0}}\delta\left[x\left(t'\right)-X\right]}{\int Dx\left(t\right)e^{-s_{\mu}\left[x\left(t\right)\right]/D_{0}}}
\end{equation}
where we defined the biased action
\begin{equation}
	\label{eq:s_lambda_def}
	s_{\mu}\left[x\left(t\right)\right]=s\left[x\left(t\right)\right]+\mu D_{0}\int_{0}^{T}x\left(t\right)dt.
\end{equation}
Comparing Eqs.~(\ref{eq:action_y}) and~(\ref{eq:P_ofDeltaX_path_integral}) with~(\ref{eq:canonical_ensemble_as_path_integral}) and~(\ref{eq:s_lambda_def}) we arrive at
\begin{equation}
	\label{eq:exact_connection2}
	P_{C}\left(\Delta X\right)=p\left(\Delta X;\mu=\frac{2C}{D_{0}T}\right).
\end{equation}

On the other hand, the canonical ensemble and the conditioned on $A$ (or micro-canonical) ensemble are equivalent and share the same relation as the one between canonical and micro-canonical ensembles in equilibrium statistical mechanics \cite{hugoprl}.
In order to write this relation explicitly, we first consider the \emph{joint} probability density $\mathcal{P}\left(X,A\right)$ of $X=x\left(t'\right)$ and $A$. It is related to the conditional probability via
\begin{equation}
	\label{eq:conditional_probdef}
	\mathcal{P}\left(X,A\right)=P\left(A\right)p\left(X|A\right),
\end{equation}
where $P\left(A\right)$ is given by the Airy distribution, see Eq.~(3) of the main text.
Next we note that, up to a normalization constant, the joint probability of $X$ and $A$ in the canonical ensemble is simply $\mathcal{P}_{\mu}\left(X,A\right) \propto e^{-\mu A}\mathcal{P}\left(X,A\right)$. As a result, the relation between the two ensembles can be written as
\begin{equation}
	\label{eq:canonical_ensemble_def}
	p\left(X;\mu\right)=\frac{F\left(X,\mu\right)}{\mathcal{N}\left(\mu\right)},
\end{equation}
where
\begin{equation}
	\label{eq:Fdef}
	F\left(X,\mu\right)=\int_{0}^{\infty}e^{-\mu A}\mathcal{P}\left(X,A\right)dA
\end{equation}
is the Laplace transform of $\mathcal{P}\left(X,A\right)$ with respect to $A$, and
\begin{equation}
	\label{eq:Ndef}
	\mathcal{N}\left(\mu\right)=\int_{0}^{\infty}e^{-\mu A}P\left(A\right)dA
\end{equation}
enforces the normalization condition $\int_{0}^{\infty}p\left(X;\mu\right)dX=1$.
Note that $\mathcal{N}\left(\mu\right)$ is the Laplace transform of $P\left(A\right)$, and it is known exactly \citep{dar, Louch}. Equations~(\ref{eq:exact_connection2})-(\ref{eq:Ndef}) provide an exact connection between the conditional probability distribution $p\left(X|A\right)$ and the distribution $P_{C}\left(\Delta X\right)$ in the Ferrari-Spohn model with a parabolic wall.

In the limit of $T\to\infty$ at fixed values of $A/T$ and $X$ (note that this limit implies a small area $A\ll D_{0}^{1/2}T^{3/2}$) the inverse Laplace transforms, that give $\mathcal{P}\left(X,A\right)$ and $P\left(A\right)$ from $F\left(X,\mu\right)$ and $\mathcal{N}\left(\mu\right)$, respectively, can be evaluated using the saddle-point approximation. This approximation is the basis of the DV formalism which we describe in Sec.~\ref{dvtheo}. As a result, the conditional distribution can be written as
\begin{equation}
	\label{eq:P_X_A_DV}
	p\left(X|A\right)
	=\frac{\mathcal{P}\left(X,A\right)}{P\left(A\right)}
	\simeq\frac{e^{\mu^{*}A}F\left(X,\mu^{*}\right)}{e^{\mu^{*}A}\mathcal{N}\left(\mu^{*}\right)}
	=\frac{F\left(X,\mu^{*}\right)}{\mathcal{N}\left(\mu^{*}\right)} ,
\end{equation}
where
$\mu^{*}=\mu^{*}\left(A\right)$ is found from the solution of the DV eigenvalue problem, see Eq.~\eqref{optlam}:
\begin{equation}
	\label{eq:lambdastar}
	\mu^{*}=-k=4D_{0}\left(\frac{\alpha_{1}T}{3A}\right)^{3}.
\end{equation}
Putting together Eqs.~(\ref{eq:exact_connection2}),~(\ref{eq:canonical_ensemble_def}),~(\ref{eq:P_X_A_DV}) and~(\ref{eq:lambdastar}), we obtain
\begin{equation}
	\label{eq:connection1}
	p\left(X|A\right)\simeq P_{C^{*}}\left(X\right),
\end{equation}
where
\begin{equation}
	\label{eq:Cstar_def}
	C^{*}=\frac{\mu^{*}D_{0}T}{2}=\frac{2\alpha_{1}^{3}D_{0}^{2}T^{4}}{27A^{3}}.
\end{equation}
Now, since
$$ \frac{C^{*}T}{\sqrt{D_{0}T}}
=\frac{2\alpha_{1}^{3}}{27}\left(\frac{D_{0}^{1/2}T^{3/2}}{A}\right)^{3}\gg1, $$
$P_{C^*}\left(X\right)$ in Eq.~(\ref{eq:connection1}) is given by the FS distribution~(\ref{eq:FS_distribution_with_ell}),
with $\ell = 2\alpha_{1}T/\left(3A\right)$ which is found by plugging Eq.~(\ref{eq:Cstar_def}) into Eq.~(\ref{eq:ell_as_function_of_C}). This indeed leads to a coincidence of Eq.~(\ref{eq:connection1})  with Eq.~(10) of the main text. To remind the reader, the coincidence occurs in the limit of $T\to\infty$ at fixed values of $A/T$ and $X$. Larger deviations in the two models (when $T$ is fixed, and we consider the large-$\Delta X$ or large-$X$ limit) behave differently, see Eqs.~(14)-(16) of the main text and Ref. \cite{SmithMeerson2019a}.

\section{The OFM and the large-$A$ tail}\label{ofm}
\renewcommand{\theequation}{D\arabic{equation}}
\setcounter{equation}{0}

The conditional probability distribution $p\left(X,t|A\gg\sqrt{D_{0}T^{3}}\right)$  is given by the ratio of the probability of the Brownian excursion realizing a specified area $A$ \emph{and} the constraint $x(t)=X$, and the probability of it realizing the area $A$ alone. Within the OFM, these probabilities are given by the path probability measures of the \textit{optimal} Brownian excursion $x(t^{\prime})$ realizing the large $A$ with and without the additional constraint $x(t)=X$. The latter is accommodated into the OFM minimization problem via an additional  Lagrange multiplier $\lambda_2$
leading to the effective Lagrangian $L\left[x\left(t^{\prime}\right),\dot{x}\left(t^{\prime}\right)\right]=\dot{x}^{2}\!/2-\lambda x-\lambda_{2}\delta\left(t^{\prime}-t\right)$, where $\delta(\dots)$ is the delta-function \cite{SmithMeerson2019a}.
There are three regimes of interest: the subcritical, the intermediate and the supercritical, separated by two third-order dynamical phase transitions, as we now describe.

\begin{figure}[h]
	\begin{tabular}{ll}
		\includegraphics[width=0.42\textwidth,clip=]{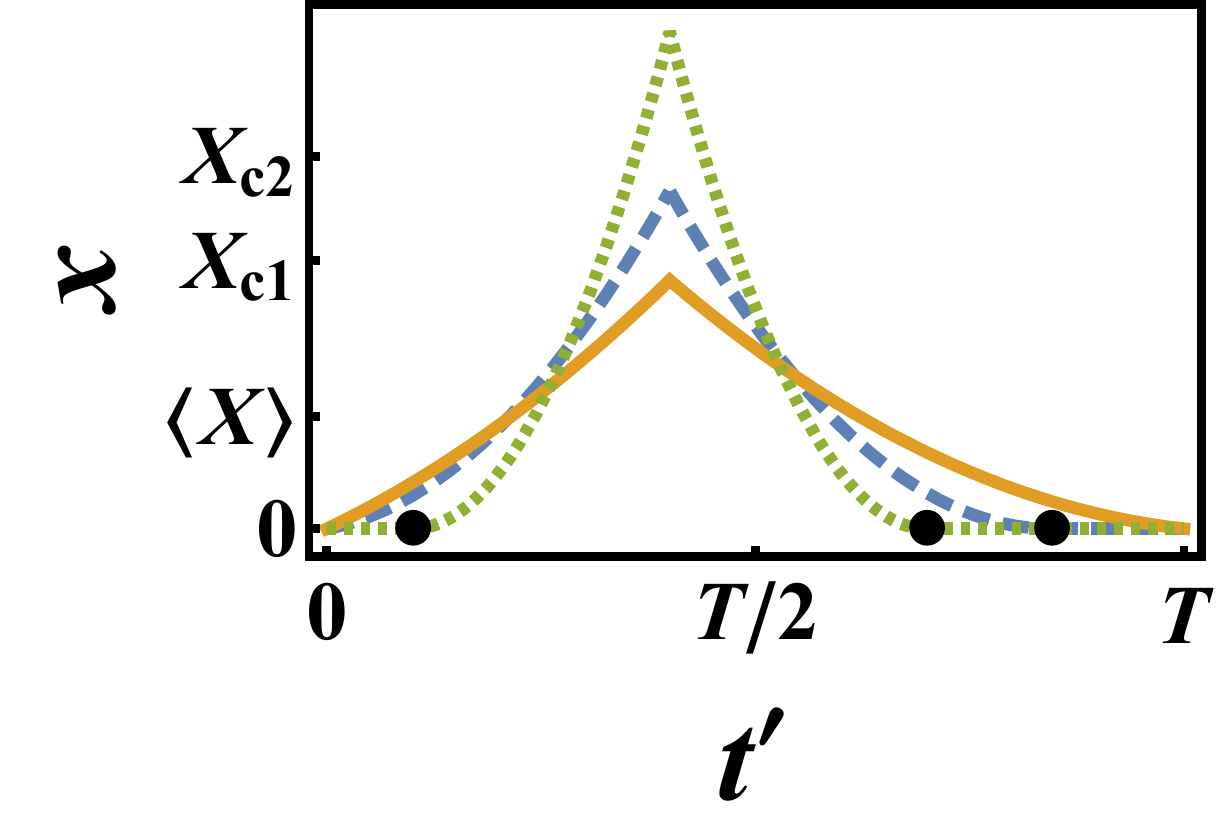}
	\end{tabular}	
	\caption{Optimal paths conditioned on $x\left(t=0.4 T\right)=X$ and the area $A$, in the subcritical (solid), intermediate (dashed) and supercritical (dotted) regimes, see Eqs.~\eqref{lagrangein}, \eqref{intermediate} and~\eqref{trajsup_gen}. The parameters are $XT/A=2.5	,3.4$ and $5$ (solid, dashed and dotted lines, respectively). Circles: the times when $x(t)$ vanishes: $t^{\prime}=\tau$ for the intermediate trajectory [see Eq.~\eqref{intermediate} and \eqref{tau}], and $|t-t^{\prime}|= 3A/\left(2X\right)$ for the supercritical trajectory.}
	\label{3regieme}		
\end{figure}

In the subcritical regime, $0<X\le X_{c1}$, where
\begin{equation}
	X_{c1}\left(t\right)=\frac{6\left(\frac{T}{2}+\left|\frac{T}{2}-t\right|\right)^{2}A}{\left(3\left|\frac{T}{2}-t\right|+\frac{T}{2}\right)T^2},
\end{equation}
the optimal trajectory is composed of two parabolic segments with a discontinuous derivative at $t^{\prime}=t$. For $0\leq t^{\prime}\leq t$ the trajectory is given by
\begin{equation}
	{\!\! x\left(t^{\prime}\right)=}
	\frac{t^{\prime}\left[X\left(T^2-3tT\right)+6at^2-t^{\prime}\left(6at-3Xt\right)\right]}{t\left(T^2-3tT+3t^2\right)}. \label{lagrangein}
\end{equation}
For  $t\leq t^{\prime}\leq T$ one should replace here $t$ and $t^{\prime}$ by $T-t$ and $T-t^{\prime}$, respectively.
This trajectory is shown by the solid line in Fig. \ref{3regieme}.
In the particular case of $t=T/2$ the trajectory is symmetric around $T/2$, see  Eq.~(12) of the main text.  The conditional probability is given by the difference in the actions [Eq.~(7) of the main text] along the trajectory \eqref{lagrangein} and the trajectory given by Eq.~(8) of the main text which is conditioned by the area $A$ alone. Here we obtain a Gaussian distribution:
\begin{equation}
	-\ln p\left(X,t|A\gg \sqrt{D_0T^3}\right)\simeq\frac{\Delta s}{D_0} =\frac{\left[X-x^*_A\left(t\right)\right]^2}{2D_0T\sigma^2(t)},
\end{equation}
where $\sigma^2\left(t\right)$ and $x^*_A\left(t\right)$ are given by Eqs.~(18) and~(8), respectively, of the main text.

When $X$ exceeds $X_{c1}$, $x(t^{\prime})$ from Eq.~(\ref{lagrangein}) would cross the origin. For $t<T/2$ ($t>T/2$) this happens first at the right (left) end of the trajectory $t^{\prime}=T$. A solution crossing the origin is not allowed, and the correct solution in this, intermediate  regime is given by the tangent construction of the calculus of one-sided variations \cite{onesided}. This solution vanishes identically past a point $\tau$ that we now determine. For concreteness, let us assume that $0<t<T/2$. (The case $T/2<t<T$ is obtained from symmetry.) In this regime, $X_{c1}\le X\le X_{c2}\equiv 3A/(2t)$,
the optimal trajectory is
\begin{equation}
	x\left(t'\right)=\frac{X}{\left(\tau-t\right)^2}\times\begin{cases}
		0 & \tau\le t'\le T,\\
		\left(\tau-t'\right)^{2} & t\le t'\le\tau,\\
		t'^2+\frac{\left(\tau-t\right)^2-t^2}{t}t' & 0\le t'\le t,
	\end{cases}\label{intermediate}
\end{equation}
where $\tau$ is given by
\begin{eqnarray}
	\frac{\tau}{t}&=&\frac{\left(2\tilde{a}-1\right)^2}{2\left[\sqrt{\left(2\tilde{a}-1\right)^3+1}+1\right]^{2/3}}\\
	&+&\frac{1}{2}\left[\sqrt{\left(2\tilde{a}-1\right)^3+1}+1\right]^{2/3}
	+\tilde{a}+\frac{1}{2}, 
\label{tau}\nonumber
\end{eqnarray}
where $\tilde{a}\equiv A/(Xt)$. This trajectory is shown by the dashed line in Fig. \ref{3regieme}.

In the supercritical regime $X\ge X_{c2}$, the trajectory (\ref{intermediate}) also becomes invalid, as it crosses the origin. Now this happens immediately: at $t^{\prime}=0$. Again, the tangent construction is needed in order to find the valid optimal trajectory. This time the correct $x(t)$ vanishes at two points along the trajectory:
\begin{equation}
	\label{trajsup_gen}
	x\left(t'\right)=\begin{cases}
		\frac{4X^{3}}{9A^{2}}\left(\frac{3A}{2X}-\left|t-t'\right|\right)^{2} & |t-t'|\leq\frac{3A}{2X},\\
		0 & \left|t-t'\right|\geq\frac{3A}{2X} .
	\end{cases}
\end{equation}
This expression, which is a simple extension of Eq.~(13) to $t \ne T/2$,  is shown by the dotted line in Fig. \ref{3regieme}.

At each of the two critical lines $X=X_{c1}\left(t\right)$ and $X=X_{c2}\left(t\right)$ a third-order dynamical phase transition occurs, corresponding to a jump in the third derivative of the large-deviation function with respect to $X$.
At $X_{c1}\leq X\leq X_{c2}$ the large-deviation function is given by
\begin{eqnarray}\nonumber
	&&-\ln p\left(X,t|A\gg \sqrt{D_0T^3}\right)\simeq\frac{\Delta s}{D_0} \\
	&&=\frac{6A^2}{D_0}\left[\frac{3\tau^2-8t\tau+6t^2}{t\tau^2\left(2\tau-3t\right)^2}-\frac{1}{T^3}\right],
\end{eqnarray}
where $\tau$ is given by \eqref{tau}, while at $X \geq X_{c2}$ the large-deviation function is given by
\begin{equation}
	-\ln p\left(X,t|A\gg \sqrt{D_0T^3}\right)\simeq\frac{\Delta s}{D_0} =\frac{A^2}{D_0}\left(\frac{8X^3}{9A^3}-\frac{1}{T^3}\right).
\end{equation}
For the particular case $t=T/2$ the two dynamical phase transitions merge into a single third-order transition, as described in the main text.
{		
	\section{Area under the square of Brownian excursion}
	\renewcommand{\theequation}{E\arabic{equation}}
	\setcounter{equation}{0}\label{square}
	
	Here we study the probability distribution $P(B,T)$ of the area under the \emph{square} of Brownian excursion,
	\begin{align}\label{B}
		B = \int_0^T dt\, x^2(t).
	\end{align}
	Dimensional analysis yields the scaling form
	$$
	P(B,T) = \frac{1}{D_0T^2} h\left(\frac{B}{D_0T^2}\right).
	$$
	For convenience we will set $D_0 = 1$ and restore the $D_0$-dependence in the final results.
	The Laplace transform  of $P(B,T) $,  $\tilde{P}(\lambda, T)=\int_0^{\infty}P\left(B,T\right)e^{-\lambda B}dB$, was derived in Ref. \cite{The} by probabilistic methods.  For completeness, we present here a simpler and more physical derivation by using path integral methods. Our main focus, however, are the small- and large-$B$ tails of $P(B,T)$ and their intrinsic connections to the DV method and the OFM, respectively.
	
	The probability distribution $P(B,T)$ is given by a sum over all the Brownian excursions on the interval $0<t<T$, conditioned by Eq. \eqref{B}. There is a subtlety here: a Brownian particle, starting at the origin, would cross it infinitely many times immediately afterwards and would not stay positive as required. This difficulty is circumvented by introducing a cutoff: assuming that $x(0)=x(T)=\epsilon>0$ and sending $\epsilon$  to zero at the end of the calculation, see \textit{e.g.} Ref. \cite{satcomt}. That is,
	\begin{align}
		\tilde{P}(\lambda, T) &= \lim_{\epsilon\to 0} \braket{e^{-\lambda\int_0^T d\tau\, x^2(\tau)}}_\epsilon ,
	\end{align}
	where $\braket{...}_\epsilon$ denotes the expectation value over realizations of Brownian motions which satisfy $x(0) = x(T) = \epsilon$ and stay positive for $0<t<T$. The expectation value is given by
\begin{eqnarray}\nonumber
		&&\braket{e^{-\lambda\int_0^T d\tau\, x^2(\tau)}}_\epsilon=\\
		&&\frac{1}{Z(\epsilon,T)} \int_{x(0)=\epsilon}^{x(T)=\epsilon}\mathcal{D}x(\tau)\, e^{-\int_0^T d\tau\, \left[ \frac{1}{2} \left(\frac{\partial x}{\partial \tau}\right)^2 + \lambda x^2(\tau) \right]}\nonumber\\
		&\times&\prod_{\tau=0}^T \theta[x(\tau)]
		= \frac{G_\lambda(\epsilon,T|\epsilon,0)}{Z(\epsilon,T)} , \label{eq:propagator}
\end{eqnarray}
	where the normalization constant $Z(\epsilon,T) = \tilde{P}(0,T)$ is the probability of a Brownian excursion unconstrained by $B$. The propagator $G_\lambda(\epsilon, T|\epsilon,0)$ is given by the quantum mechanical expectation value $\braket{\epsilon|e^{-\hat{H}^{(\lambda)} T}|\epsilon}$, with the Hamiltonian $\hat{H}^{(\lambda)} = -(1/2)\, \partial^2_{x} + \lambda V(x)$, where the potential is given by $V(x>0) = x^2$ and $V(x\leq 0)=\infty$.  The normalization constant $Z(\epsilon,T)$ can be obtained either by solving the diffusion equation with absorbing boundary condition at the origin \cite{Redner}, or by applying the Feynman-Kac formula \cite{satyaprl}, which reduces the problem to finding the propagator $G_0\left(\epsilon, T | \epsilon, 0\right) = \braket{\epsilon | e^{-\hat{H}_1} | \epsilon}$, where $\hat{H}_1 = -(1/2)\, \partial^2_{x} + V(x)$ with $V(x>0) = 0$ and $V(x\leq 0)=\infty$. Both methods yield
	\begin{equation}\label{Z}
		Z(\epsilon,T) = \frac{1}{\sqrt{2\pi T}}\left(1 - e^{-\frac{2\epsilon^2}{T}}\right).
	\end{equation}
	Let us calculate the propagator $G_\lambda(\epsilon, T|\epsilon,0)$.  $\hat{H}^{(\lambda)}$ is the Hamiltonian of a harmonic oscillator, where $\hbar = m=1$ and $\omega^2 = 2\lambda>0$. Because of the absorbing boundary condition at $x=0$, only the odd eigenfunctions (the Gauss-Hermite functions \cite{landau}) are present.  The spectrum is given by $E_{2k+1} = \sqrt{2\lambda}\left(2k + 3/2 \right)$. The propagator $G_\lambda$ can be expanded in this basis:
	\begin{eqnarray}
		G_\lambda(\epsilon,t|\epsilon,0) &=& \braket{x=\epsilon | e^{-\hat{H}^{(\lambda)} T} | x=\epsilon}\\
		&=& \sum_{k=0}^\infty |\psi_{2k+1}(\epsilon)|^2 e^{-\sqrt{2\lambda }\left(2k + \frac32\right) T} \label{eq:sum} \nonumber,
	\end{eqnarray}
	where the Gauss-Hermite functions, normalized to unity over the region $x\in [0, \infty)$, are  \cite{landau}
	\begin{align}\label{eq:GaussHermiteFunction}
		\psi_{2k+1}(\epsilon) &= \frac{\sqrt{2}}{\sqrt{2^{2k+1}(2k+1)!}} \left(\frac{\omega}{\pi}\right)^{1/4} e^{-\frac{\omega\epsilon^2}{2}} H_{2k+1}\left(\sqrt{\omega}\epsilon\right),
	\end{align}
	where $H_{2k+1}(\dots)$ are the Hermite polynomials, and $\omega^2 = 2\lambda$. Before we send $\epsilon$ to zero, we evaluate Eq.~(\ref{eq:GaussHermiteFunction}) at small $\epsilon$. The small-argument asymptotic of $H_{2k+1}$ \citep{HermitePoly} can be written as
	\begin{align}
		H_{2k+1}(\epsilon) \simeq \frac{(-1)^{k}(2k+2)!}{(k+1)!}\epsilon .
	\end{align}
Plugging this asymptotic into the Gauss-Hermite functions \eqref{eq:GaussHermiteFunction}, we obtain
	\begin{align}\label{eq:GaussHermiteFunction_firstOrder}
		\psi_{2k+1}(\epsilon) &\simeq
		\frac{(-1)^{k}\sqrt{(2k+1)!}}{2^{k-1}k!} \left(\frac{\omega}{\pi}\right)^{1/4} \sqrt{\omega}\epsilon .
	\end{align}
	Using this expression in \eqref{eq:sum},  we can perform the summation exactly:
	\begin{eqnarray}\nonumber
		G_\lambda(\epsilon, T|\epsilon, 0) &\simeq& \epsilon^2 \sum_{k=0}^\infty \frac{(2k+1)!\, e^{-\sqrt{8\lambda}Tk}}{2^{2k-2}(k!)^2} e^{-\sqrt{\frac {9\lambda}{2}}T} \left[\frac{(2\lambda)^3}{\pi^2}\right]^{1/4}\\
		&=& \epsilon^2\frac{4e^{-\sqrt{\frac{9\lambda}{2}}T}}{\left( 1 - e^{-\sqrt{8\lambda}T}\right)^{3/2}}
		  \left[\frac{(2\lambda)^3}{\pi^2}\right]^{1/4} .
	\end{eqnarray}
	Using Eq.~(\ref{Z}), we can now calculate the propagator $\tilde{P}(\lambda, T) = \lim_{\epsilon\to 0} \left[G_\lambda(\epsilon, T)/Z(\epsilon,T)\right]$. Restoring the $D_0$-dependence, we finally obtain
	\begin{align}
		\tilde{P}(\lambda, T) &= \left[ \frac{\sqrt{2D_0\lambda}\, T}{\sinh\left(\sqrt{2D_0\lambda}\,T\right)} \right]^{3/2} , \label{eq:finalPropagator}
	\end{align}
	in agreement with Ref. \cite{The}.  To our knowledge, the inverse Laplace transform of Eq.~(\ref{eq:finalPropagator}) is unknown in analytical form. We performed the inversion numerically,  see Fig.~\ref{squaredInverseLaplace}, using a multi-precision inverse Laplace transform algorithm \cite{multiprecision}, realized in ``Mathematica".
\begin{figure}[h]
			\includegraphics[width=0.42\textwidth,clip=]{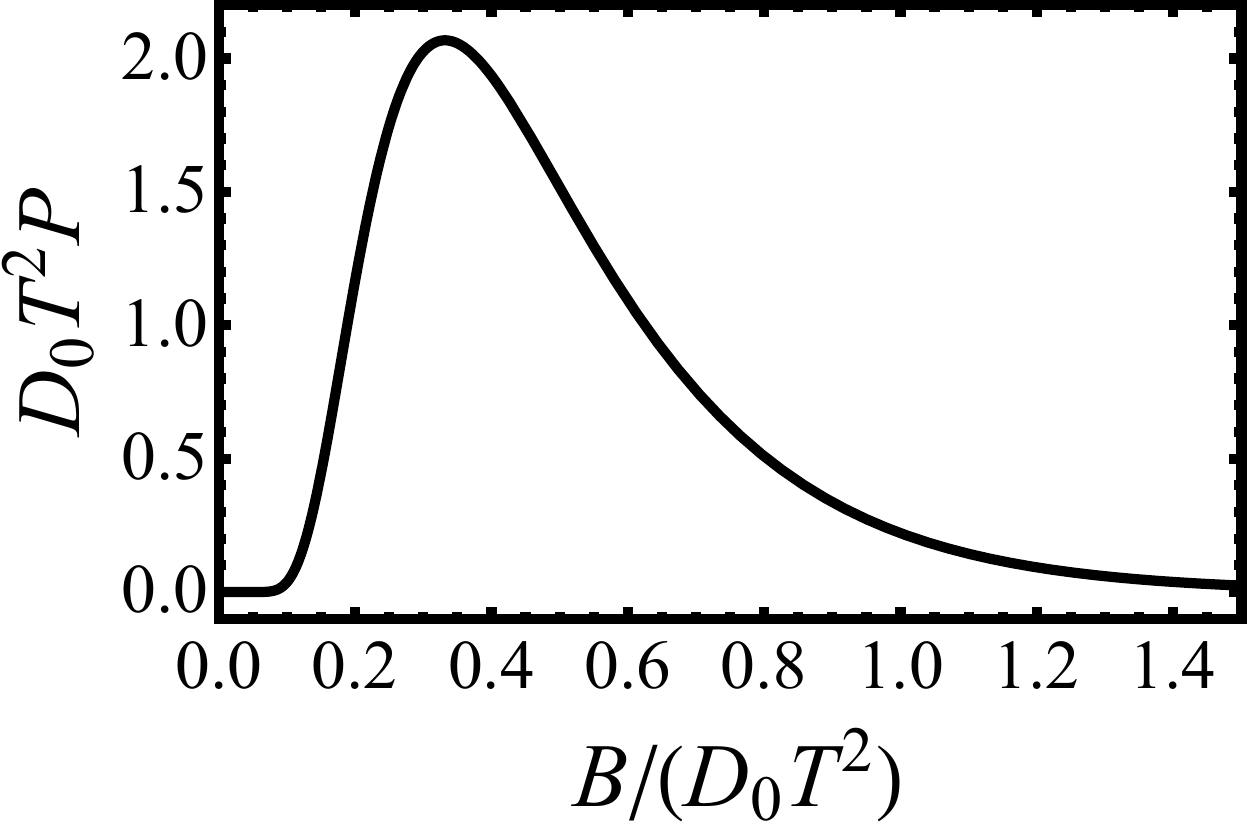}
\caption{The rescaled probability distribution $P\left(B,T\right)$ of the area $B$ under the square of a Brownian excursion. }
		\label{squaredInverseLaplace}		
\end{figure}
	
	We shall now derive the small- and large-$B$ tails of the distribution $P(B,T)$. Instead of extracting them from the exact Laplace transform (\ref{eq:finalPropagator}), we will directly employ the DV formalism for small $B$ and the OFM for large $B$. As we will see, these calculations turn out to be quite simple.

\vspace{-0.1 cm}
\subsection{Small-$B$ tail of $P(B,T)$}
\vspace{-0.1 cm}
By virtue of Eq.~\eqref{Quantum} for $\hat{H}^{(k)} \equiv \hat{H}^{(-\lambda)}$, we obtain
	\begin{align}
		\tilde{I}(k) &= -\sqrt{\frac{-9k}{2}} ,
	\end{align}
where $k < 0$. The Legendre-Fenchel transform \eqref{GE} yields the DV rate function
	\begin{align}
		I(b) = \frac{9}{8b}
	\end{align}
where $b \equiv B/T$. This leads to
	\begin{align}
		-\ln P(B \ll D_0T^2) &\simeq \frac{9D_0T^2}{8B}, \label{squaredDV}
	\end{align}
in agreement with Ref. \cite{cso}.
\vspace{-0.1 cm}
\subsection{Large-$B$ tail of $P(B,T)$}
\vspace{-0.1 cm}
	
The constrained Lagrangian of the OFM is now $L\left[x(t),\dot{x}(t)\right] = \dot{x}^2/2 - \lambda x^2$. The optimal trajectory is
	\begin{align}
		x(t) = \sqrt{\frac{2B}{T}}\sin\left(\frac{\pi t}{T}\right),
	\end{align}
	where we set $\lambda = \pi^2/(2T^2)$ to obey the constraint $B=\int_0^T x^2(t) dt$. Calculating the action from Eq.~(7), we finally obtain
	\begin{align}
		-\ln P(B \gg D_0T^2) &\simeq \frac{\pi^2 B}{2D_0T^2}, \label{squaredOFM}
	\end{align}		
in agreement with Ref. \cite{cso}.


\begin{thebibliography} {99}
\bibitem{dif}
The Brownian motion can be formally defined by the equation $\dot{x}=\sqrt{D_0}\,\xi(t)$, where $\xi(t)$ is a delta-correlated zero-mean Gaussian noise, $\braket{\xi\left(t\right)\xi\left(t^{\prime}\right)}=\delta\left(t-t^{\prime}\right)$.
\bibitem{inaccess}
Methods like Doob's $h$-transform \cite{doob} 
are hard to use due to the conditioning on both $X$ and $A$ \cite{conditionedsatya1,conditionedsatya2}. An alternative method, proposed   for a generalized Brownian bridge \cite{genbri,mazollo,mazollo2},   could not be extended to excursions, as it is restricted to Gaussian processes.
\end{thebibliography}
\end{document}